\documentclass[floatfix,
    10pt,
    twocolumn,
    aps,
    pra,
    amsmath,
    amssymb,
    mathrsfs,
    superscriptaddress,
]{revtex4-2}
\usepackage[utf8]{inputenc}
\usepackage{graphicx}
\usepackage{dcolumn}
\usepackage{bm}
\usepackage{braket}
\usepackage{epsfig}
\usepackage{color}
\usepackage{orcidlink}
\usepackage{hyperref}
\usepackage{siunitx}

\hypersetup{colorlinks=true,linkcolor=blue,citecolor=blue}

\begin{document}
\title{Infrared absorption spectroscopy of a single polyatomic molecular ion}
\author{Zhenlin Wu \orcidlink{0000-0002-8188-7701}}
\affiliation{Institut f\"ur Experimentalphysik, Universit\"at Innsbruck, Technikerstraße 25/4, 6020 Innsbruck, Austria}

\author{Tim Duka \orcidlink{0009-0002-5850-6154}}
\affiliation{Institut f\"ur Experimentalphysik, Universit\"at Innsbruck, Technikerstraße 25/4, 6020 Innsbruck, Austria}

\author{Mariano Isaza-Monsalve \orcidlink{0009-0006-6996-7916}}
\affiliation{Institut f\"ur Experimentalphysik, Universit\"at Innsbruck, Technikerstraße 25/4, 6020 Innsbruck, Austria}

\author{Miriam Kautzky \orcidlink{0009-0009-9819-6702}}
\affiliation{Institut f\"ur Experimentalphysik, Universit\"at Innsbruck, Technikerstraße 25/4, 6020 Innsbruck, Austria}

\author{Vojtěch Švarc \orcidlink{0000-0001-5535-8243}}
\affiliation{Department of Optics, Palacký University, 17. listopadu 12, 771 46 Olomouc, Czech Republic}

\author{Andrea Turci \orcidlink{0009-0004-1323-6087}}
\affiliation{Institut f\"ur Experimentalphysik, Universit\"at Innsbruck, Technikerstraße 25/4, 6020 Innsbruck, Austria}

\author{René Nardi \orcidlink{0009-0002-7533-6126}}
\affiliation{Institut f\"ur Experimentalphysik, Universit\"at Innsbruck, Technikerstraße 25/4, 6020 Innsbruck, Austria}

\author{Marcin Gronowski \orcidlink{0000-0002-7547-4548}} 
\affiliation{Faculty of Physics, University of Warsaw, Pasteura 5, 02-093 Warsaw, Poland}

\author{Micha{\l} Tomza \orcidlink{0000-0003-1792-8043}} 
\affiliation{Faculty of Physics, University of Warsaw, Pasteura 5, 02-093 Warsaw, Poland}

\author{Brandon J. Furey \orcidlink{0000-0001-7535-1874}}
\affiliation{Institut f\"ur Experimentalphysik, Universit\"at Innsbruck, Technikerstraße 25/4, 6020 Innsbruck, Austria}

\author{Philipp Schindler \orcidlink{0000-0002-9461-9650}}
\email{philipp.schindler@uibk.ac.at}
\affiliation{Institut f\"ur Experimentalphysik, Universit\"at Innsbruck, Technikerstraße 25/4, 6020 Innsbruck, Austria}
\affiliation{Department of Optics, Palacký University, 17. listopadu 12, 771 46 Olomouc, Czech Republic}

\date{\today}

\begin{abstract}
Absorption spectroscopy is a fundamental tool for probing molecular structure~\cite{hollasModernSpectroscopy2004}. However, performing absorption spectroscopy on individual molecules is challenging due to the low signal-to-noise ratio~\cite{gaidukRoomTemperatureDetectionSingle2010, whittakerAbsorptionSpectroscopyUltimate2017}. Here, we report on a non-destructive absorption spectroscopy on a mid-infrared vibrational transition in a single molecular ion that is co-trapped with an atomic ion. The absorption of a single photon is detected via the momentum transfer from the absorbed photon onto the molecule. This recoil signal is amplified using a non-classical state of motion of the two-ion crystal and subsequently read out via the atomic ion~\cite{hempelEntanglementenhancedDetectionSinglephoton2013}. We characterize the recoil detection method and use it to investigate the interaction between femtosecond laser pulses and the O--H stretching vibration in individual CaOH$^+$ molecular ions. Furthermore, we present the spectrum obtained for the vibrational transition with single-photon sensitivity. This method can provide a way of performing non-destructive state detection for the measurement and preparation of the quantum state of a wide range of molecular species. 
\end{abstract}

\maketitle

\section{Introduction}
\begin{figure}
    \centering
    \includegraphics[width=0.9\linewidth]{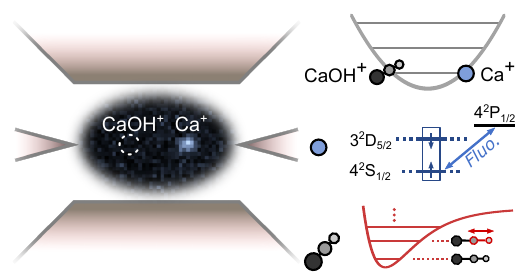}
    \caption{Schematic of the system of a trapped two-ion mixed-species crystal for performing molecular spectroscopy. The system is considered here to possess three degrees of freedom: a harmonic oscillator describing the in-phase motion of the crystal, a two-level system in the atomic ion for quantum logic operations, and the intramolecular vibration.}
    \label{fig:fig1}
\end{figure}

Absorption spectroscopy is one of the most fundamental methods to investigate light-matter interaction and has long been accepted as a standard method~\cite{hollasModernSpectroscopy2004}. Usually, the fraction of the transmitted light is measured, which reveals quantities such as the absolute absorption cross section~\cite{berdenCavityRingdownSpectroscopy2000,banwellFundamentalsMolecularSpectroscopy2017}. Measuring the fraction of absorbed light becomes difficult when investigating a single atom or molecule, as fluctuations and the inherent quantum noise of light often dominates the signal~\cite{gaidukRoomTemperatureDetectionSingle2010, whittakerAbsorptionSpectroscopyUltimate2017}.
Nevertheless, there have been several demonstrations of detecting the absorption of light in a single atom or molecule with visible light~\cite{winelandAbsorptionSpectroscopyLimit1987,celebranoSinglemoleculeImagingOptical2011,streedAbsorptionImagingSingle2012}. These experiments are performed with a large number of photons being absorbed by the atom or molecule to produce a significant absorption signal and require efficient and low-noise photon detectors which are not available for a large part of the electromagnetic spectrum. 

For molecular ions which repel each other, it is difficult to obtain a dense sample for performing efficient absorption spectroscopy and often requires techniques that are applicable to single molecules. Thus, it is common to investigate molecular ions through the secondary actions that occur when the molecule absorbs light, such as photodissociation~\cite{okumuraVibrationalPredissociationSpectroscopy1985,koelemeijVibrationalSpectroscopyMathrm2007,lohPrecisionSpectroscopyPolarized2013}, charge transfer~\cite{germannObservationElectricdipoleforbiddenInfrared2014}, fragmentation~\cite{schmidLeakOutSpectroscopyUniversal2022}, and inelastic collisions \cite{calvinSingleMoleculeInfrared2023,calvinRotationallyResolvedSpectroscopy2025}. In contrast to these methods, which often perturb the molecular state or even destroy the molecular ions, recent efforts have been made to utilize quantum logic spectroscopy for detecting the state of the molecular ion via a co-trapped atomic ion in a non-destructive fashion~\cite{wolfNondestructiveStateDetection2016,sinhalQuantumnondemolitionStateDetection2020, chouPreparationCoherentManipulation2017,holzapfelQuantumControlSingle2025}. Quantum logic spectroscopy enables spectroscopy on intramolecular transitions by monitoring the state of the system and its changes due to interaction with light~\cite{chouFrequencycombSpectroscopyPure2020,sinhalQuantumnondemolitionStateDetection2020}. To date, these experiments are limited to molecular species with relatively simple internal structure and are predominantly applied to diatomic molecules.

We utilize the same principle of quantum logic spectroscopy, probing the light-molecule interaction via a co-trapped atom, but we choose to detect photon absorption events directly via the recoil that a single absorbed photon exerts on the molecule. This detection technique is known as recoil spectroscopy and has been applied to study atomic transitions~\cite{hempelEntanglementenhancedDetectionSinglephoton2013,wanPrecisionSpectroscopyPhotonrecoil2014}. Similar to traditional quantum logic spectroscopy experiments, the molecular ion is co-trapped with an atomic ion on which the recoil can be read out, as sketched in Figure~\ref{fig:fig1}. The recoil is transferred onto the atom via the Coulomb interaction, which couples the external motion of both trapped ions. The quantum state of the atom's motion can then be mapped onto its electronic states and read out using quantum information processing techniques~\cite{winelandExperimentalIssuesCoherent1998}.

Usually, the recoil of a single photon is so small that its effect on the motion of the atom cannot be measured straightforwardly for reasonable experimental parameters. However, it is possible to amplify the signal of such a recoil by exploiting non-classical states of motion of both ions. For instance, it has been shown that preparing the motion in a so-called ``Schr\"odinger cat" state enables the detection of photon absorption ~\cite{hempelEntanglementenhancedDetectionSinglephoton2013}.  A sketch of the measurement procedure is shown in Figure~\ref{fig:fig2}. To date, this method has only been demonstrated on atomic transitions where the collective recoil of multiple photon absorption and emission events was measured~\cite{hempelEntanglementenhancedDetectionSinglephoton2013}. In this work, we demonstrate the first implementation of this method for detecting single photon absorption on a molecular ion. 

\section{Photon absorption detection}

\begin{figure}
    \centering
    \includegraphics[width=0.85\linewidth]{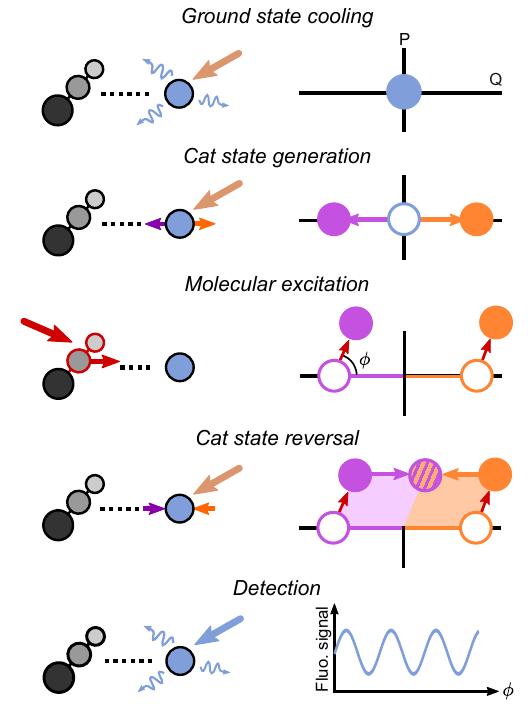}
    \caption{Sequence diagram of the cat state spectroscopy for detecting the photon absorption recoil on the molecule (left panels) and the evolution of the motional wavepacket of the ion crystal in phase space (right panels). The procedure consists of the following steps: initial ground state cooling of the ion motion, generation of the recoil sensitive cat state, excitation of the molecular transition, reversal of the cat state generation operation, and the detection of the photon absorption recoil on the atomic ion. This process generates entanglement between the atomic qubit and the motion of the ion crystal, such that signal of single photon absorption recoil is amplified and transformed into a geometric phase illustrated by the shaded area. This is then mapped to the state of the atomic ion which can be detected via fluorescence.}
    \label{fig:fig2}
\end{figure}

To illustrate the absorption detection method, we consider an atomic and a molecular ion co-trapped in a Paul trap, where we aim to detect photon absorption on the molecular ion. 
Regardless of the nature of the molecular transition, the excitation is detected based on its influence on the motion of the two ions. We can thus describe the detection method with the internal state of the atomic ion and the combined motional state of the ion crystal. 
The internal state of the atom is modeled by a two-level system with energy eigenstates $\lvert\uparrow \rangle$ and $\lvert\downarrow \rangle$, which can be manipulated and read out by applying laser pulses~\cite{winelandExperimentalIssuesCoherent1998}. 
It is convenient to describe the motional states of the two ions by their collective motional modes that arise due to the Coulomb interaction. In our case, the relevant collective motion is given by a single-mode harmonic oscillator with its general state in the Fock state basis $\Psi_\text{z} = \sum_n c_n \lvert n \rangle$. 
In particular, we target the axial in-phase mode of the mixed-species two-ion crystal~\cite{homeQuantumScienceMetrology2013}. 

We will first introduce the dynamics of a single photon absorption event in this system. Momentum conservation dictates that the absorption of a single photon not only excites a molecular transition but also transfers momentum from the photon onto the molecule. In an ion trap, this momentum transfer can be described by applying a displacement operator $\hat{D}(i \eta_\text{m})$~\cite{ozeriErrorsTrappedionQuantum2007}. The magnitude of this displacement is given by the corresponding Lamb-Dicke parameter of the molecular transition
\begin{equation}
    \label{eq:ld param}
    \eta_\text{m} = \sqrt{\frac{\hbar}{2 M_\text{m} \omega_z}} k_z^\text{m} e_z^\text{m},
\end{equation}
where $M_\text{m}$ is the molecular mass, $\omega_z$ is the motional frequency of the in-phase mode, $k_z^\text{m}$ is the wavevector component of the light exciting the molecular transition along the motional axis, and $e_z^\text{m}$ is the participation factor of the molecular ion in the in-phase mode~\cite{homeQuantumScienceMetrology2013}. 

We consider first the case where the motion is initially prepared in its ground state, which is displaced into a coherent state with amplitude $\eta_\text{m}$ due to the recoil.
For molecular transitions in the optical and infrared regime, this displacement magnitude is usually on the order of $10^{-2}$. One can thus approximate the displaced motional state with $\Psi_\text{z} = \lvert 0 \rangle + i \eta_\text{m} e^{-i \omega_z t} \lvert 1 \rangle$ as $\eta_\text{m} \ll 1$, where $t$ is the timing between the photon absorption event and the detection. 
The most efficient method for detecting this displacement directly is a phase-sensitive measurement which maps the motional state onto the basis states of the atom followed by a measurement in the atomic $\hat{\sigma}_y$ basis~\cite{hempelEntanglementenhancedDetectionSinglephoton2013}. When varying the timing of the photon absorption event, this measurement yields a sinusoidal signal with peak-to-peak contrast $\mathcal{S}_\text{direct}(\eta_\text{m}) = 2\eta_\text{m}$. In practice, this signal is too small to be measured with confidence due to fundamental and technical noise of the experiment such as quantum projection noise and imperfect motional ground state cooling. 

A more efficient way of detecting the displacement is to prepare the motional state of the ion crystal in a non-classical state before the absorption event, as described in Figure~\ref{fig:fig2}. This is done by applying a bichromatic light field to the atomic and motional ground state that causes a state-dependent force on the ions (see Methods~\hyperref[meth:cat state engineer]{B}). It generates entanglement between the in-phase motion of the crystal and the atomic qubit state. The resulting state is known as a cat state:
\begin{equation}
    \label{eq:cat state}
    \Psi_{\text{cat}} = \frac{1}{\sqrt{2}} ( \lvert + \rangle \lvert\alpha\rangle + \lvert - \rangle \lvert-\alpha\rangle ),
\end{equation}
where $\lvert \pm \rangle = (\lvert\uparrow \rangle \pm \lvert\downarrow \rangle)/\sqrt{2}$ and $\lvert \pm \alpha\rangle$ is a motional coherent state of complex amplitude $\pm \alpha$~\cite{zhangCoherentStatesTheory1990}. While $\lvert \alpha \rvert$ is referred to as the amplitude of the cat state, the argument of $\alpha$ can be interpreted as the phase of the cat state, which is controlled by the relative phase $\phi_-$ between the two frequency components of the bichromatic laser field.

A single photon absorption event acts then as a displacement on this cat state. This displacement can be mapped onto the electronic state of the atom by inverting the dynamics of the cat state generation process, which is implemented by altering the phases of bichromatic light. The state after the displacement and the inverse cat state dynamics becomes
\begin{equation}
    \begin{aligned}
        \Psi' & = \frac{1}{\sqrt{2}} (\lvert + \rangle e^{i\Phi} + \lvert - \rangle e^{-i\Phi} ) \otimes |i\eta_\text{m}\rangle  \\
        & = \left(\lvert \uparrow \rangle \cos{\Phi} + \lvert \downarrow \rangle \sin{\Phi} \right) \otimes |i\eta_\text{m}\rangle,
    \end{aligned}
    \label{eq:final state}
\end{equation}
where the rotation angle of the atomic electronic state $\Phi=2\eta_\text{m} \text{Re}(\alpha)$ reflects the amplification of the recoil. The amplification depends on $\text{Re}(\alpha)$, which is determined by the phase difference between the cat state and the displacement operator from the photon absorption recoil. The phase of the cat state is controlled by $\phi_-$ whereas the phase of the displacement operator is determined by the relative timing of the photon absorption event and the generation of the cat state. In our experiment, we chose to vary $\text{Re}(\alpha)$ between $-|\alpha|$ and $|\alpha|$ by changing $\phi_-$ while keeping the timing of the photon absorption fixed (see Methods~\hyperref[meth:imperfections]{C}). 

Overall, an experimental sequence consists of cat state generation, photon absorption, and the reversal of cat state generation. After such a sequence, the displacement from the absorption event can be related to an atomic excitation to $\lvert\downarrow\rangle$ which can be detected with a maximum probability of $\sin{(2\eta_\text{m} |\alpha|)}^2$.
In the regime of $\eta_\text{m} \text{Re}(\alpha) \ll 1$, it is advantageous to evaluate the atomic excitation in the $\hat{\sigma}_y$ basis by adding an extra $\hat{R}_x(\pi/2)$ operation on the atom before detection. We define the measured signal in this case to be the peak-to-peak amplitude in the atomic excitation as we vary $\phi_-$: 
\begin{equation}
    \label{eq:signal model}
    \mathcal{S}(\eta_\text{m}) = \sin{(4\eta_\text{m} |\alpha|)}, 
\end{equation}
which is the expected signal from a noiseless experiment.

\begin{figure}
    \centering
    \includegraphics[width=0.9\linewidth]{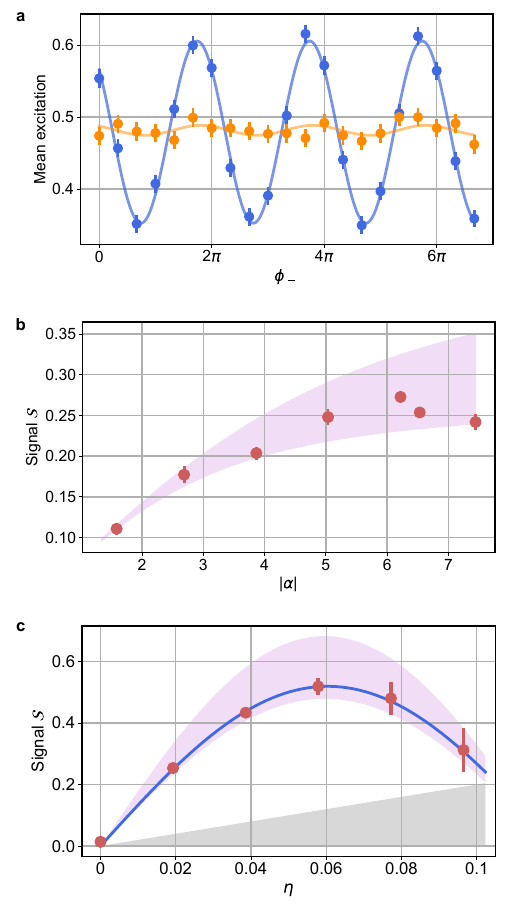}
    \caption{Characterization of the cat state spectroscopy on an electric kick-induced displacement. (a)~Measured atomic excitation signal as a function of the bichromatic light phase $\phi_-$ with $|\alpha|=6.5(7)$ on the displacement of $\eta_k = 0.0193(2)$ for $n_\text{kick}=1$ (blue) as well as the background signal for $n_\text{kick}=0$ (orange) due to quantum projection noise. Error bars are one standard deviation due to quantum projection noise. (b)~Variation of the signal with respect to cat state amplitude $|\alpha|$ for a fixed displacement, as compared to theory based on the measured atomic and motional coherence (purple, see Methods~\hyperref[meth:imperfections]{C})). Error bars are one standard deviation from the least-squares fit covariance. (c)~The signal measured with $|\alpha|=6.5(7)$ for electric displacements with various magnitudes $\eta$, as compared to theory based on the measured atomic and motional coherence in purple and a simple heuristic model in blue. The theoretically achievable contrast from a direct measurement is indicated in the gray shaded region.}
    \label{fig:fig3}
\end{figure}

\begin{figure*}
    \centering
    \includegraphics[width=\linewidth]{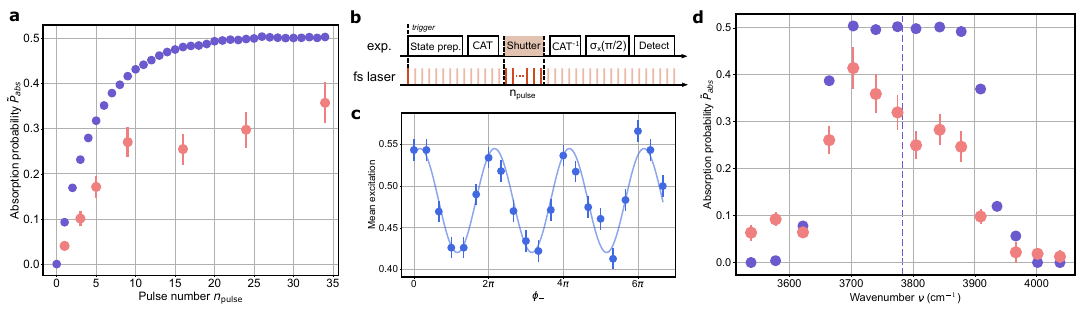}
    \caption{Absorption spectroscopy of the O--H stretching mode of CaOH$^+$. (a)~The effective photon absorption probability from the cat state spectroscopy measured at $\nu=3703$~\unit{\per\centi\metre} with different laser pulse number $n_\text{pulse}$ (red) and the results from a Monte-Carlo simulation (purple). (b)~Schematic of the pulse sequence triggered on the femtosecond laser pulse train. (c)~An example of the detected atomic excitation variation ($n_\text{pulse}=34$) in the measurement from which the peak-to-peak amplitude $\mathcal{S}$ is obtained. (d)~Absorption spectrum measured with cat state spectroscopy (red dots), along with the expected spectrum from simulation (purple dots). The value for the transition frequency from ab initio theory is indicated (purple line).}
    \label{fig:fig4}
\end{figure*}

We implemented this technique experimentally on a two-ion crystal of one $^{40}$Ca$^+$ ion and one $^{40}$CaOH$^+$ ion, which is generated by leaking in water vapor~\cite{okadaAccelerationChemicalReaction2003,wuPhotodissociationSpectraSingle2024a}, as described in Methods~\hyperref[meth:molecule generation]{A}. In order to analyze the effect of imperfections in the experiment on this method (mainly due to atomic and motional decoherence), we emulate the absorption event by a displacement caused by an electric force on the ion crystal. To implement a displacement with a controlled magnitude, we apply a series of $n_{\text{kick}}$ short electric kick pulses. They are temporally separated by the oscillation period of the ion crystal's in-phase motion such that they add up coherently to a displacement $\hat{D}(\eta=n_{\text{kick}} \eta_k)$ on the motional state of the crystal. Here, $\eta_k$ is the displacement magnitude from a single kick pulse which is set to 0.0193(2) to be comparable with the expected recoil of a mid-infrared photon absorption event.

The sensitivity of the generated cat states can then be tested with these electric impulses by varying the phase $\phi_-$ and measuring the atomic excitation in the $\hat{\sigma}_y$ basis. This yields a sinusoidal signal as illustrated in Figure~\ref{fig:fig3}(a), which shows the measured signals for a cat state amplitude of $|\alpha|=6.5(7)$ when applying a single kick and applying no kick. The signal is smaller than what is expected from Equation~\ref{eq:signal model} due to experimental imperfections. We quantify the effect of the imperfections on the achievable signal for various cat state amplitudes $|\alpha|$ as shown in Figure~\ref{fig:fig3}(b). In a noiseless experiment, the signal should increase with larger $|\alpha|$ according to Equation~\ref{eq:signal model}, but the actual signal measured at large $|\alpha|$ is limited by atomic and motional decoherence. 
This behavior is expected as the experiment duration and the sensitivity to noise increase with the amplitude of the cat state, as shown in simulations of the experiment detailed in Methods~\hyperref[meth:imperfections]{C}.
In order to reach the maximum detection efficiency, we operate our spectroscopy measurements with $|\alpha|=6.5(7)$. 
We model experimental imperfections by introducing a correction factor $\mathcal{S}_\text{max}$ that reduces the measured signal expected from Equation~\ref{eq:signal model}: 
\begin{equation}
    \label{eq:exp signal model}
    \mathcal{S}(\eta) = \mathcal{S}_{\text{max}} \sin{(4\eta |\alpha|)}
\end{equation}
where Figure~\ref{fig:fig3}(c) shows results for $\mathcal{S}_{\text{max}} = 0.52(3)$ which will be used in the following analysis. 

The measured contrast can be directly converted to a value that is relevant for absorption spectroscopy: the photon absorption probability. For probing typical molecular vibrational transitions, we assume that the molecule absorbs no more than one photon throughout the sequence and that no spontaneous decay occurs during this period. With this, we can extract an effective photon absorption probability from the measured signal $\mathcal{S}$ and the signal expected from absorbing a photon at the probe wavelength $\mathcal{S}(\eta_\text{m})$ given by Equation~\ref{eq:exp signal model}:
\begin{equation}
    \label{eq:absorption prob}
    \tilde{p}_\text{abs}=\mathcal{S}/\mathcal{S}(\eta_\text{m}).
\end{equation}
In the following, we demonstrate vibrational spectroscopy on the molecule using this effective single-photon absorption probability as a figure of merit.

\section{Molecular spectroscopy}
We now characterize a molecular vibrational transition in CaOH$^+$ using the cat state detection method. In particular, we address the vibrational transition between the ground and first excited state of the O--H stretching mode. This transition is predicted at $\nu_0=3783$~\unit{\per\centi\metre} and should be infrared active, based on ab initio calculations (details in Methods~\hyperref[meth:quantum-chemical calculations]{D}), which corresponds to $\eta_\textrm{m} \approx 0.03$.
We expect that the equilibrium state of this internal vibration is effectively the ground state, as the molecule should thermalize to its ambient temperature~\cite{liuQuantumStateTracking2024}. Repeated application of the spectroscopy sequence should not change this equilibrium distribution as the time between excitation events is long compared to the radiative lifetime of the excited state, estimated with 2.8~ms. 
Owing to the rotational substructure within the target vibrational transition, the excitation dynamics are also influenced by the thermal distribution of the rotational states. However, the molecule is excited with a femtosecond laser which has a bandwidth that cannot resolve the individual rotational lines. Therefore, our modeling of the molecule-laser interaction does not consider the rotational substructure and consists of only the two lowest energy vibrational states.

To achieve a significant excitation probability, we apply multiple laser pulses in a train consisting of up to 34 pulses. Similar to the electric impulses described above, the timing of the laser pulses is synchronized with the external motion of the ion crystal. We fulfill this by setting the in-phase motional frequency of the ion crystal to $\omega_z/2\pi=4 \cdot f_\text{rep}$, where $f_\text{rep}=100$~kHz is the repetition rate of the laser pulses, as illustrated in Figure~\ref{fig:fig4}(b). 

At a laser center frequency of $\nu=3703.3(2)$~\unit{\per\centi\metre}, we achieve a photon absorption signal of $\mathcal{S}=$0.12(1)  after applying $n_\text{pulse}=34$ pulses, shown in Figure~\ref{fig:fig4}(c). 
This cat state spectroscopy signal can then be converted into an effective photon absorption probability. As illustrated in Figure~\ref{fig:fig4}(a), we measured the photon absorption probability at the same laser center frequency as a function of the number of laser pulses.
This allows us to compare the experimental data to a simple theoretical model with no free parameters, which predicts the absorption probability to saturate to 0.5 due to the random orientation of the molecule and dephasing (see Methods~\hyperref[meth:molecular excitation dynamics]{E}).

A spectrum of the O-H stretching transition measured with this technique is shown in Figure~\ref{fig:fig4}(d). These measurements are performed by varying the center frequency of the laser between 3600~\unit{\per\centi\metre} and 4000~\unit{\per\centi\metre} and applying a train of 34 laser pulses. It is observed that the center of the absorption peak is close to the value of $\nu_0=3783$~\unit{\per\centi\metre} from ab initio calculations and that the width of the simulated absorption peak matches the measured spectrum. However, the model fails to explain the absolute magnitude of the measured effective photon absorption probability. 

The comparison of the experiment with the model indicates that the model captures the qualitative features but does not fully explain the observed dynamics and spectral structure.
We plan to advance both the experimental and theoretical analysis further to develop a more accurate model for the obtained photon absorption spectrum. 
In particular, we aim to increase the spectroscopy laser intensity to reach appreciable excitation probability with a single laser pulse. 
This will facilitate a more sophisticated theoretical modeling of the coherence of the molecule-light interaction and will enable precise measurement of the transition properties.  

In the presented experiments, we chose to use broadband femtosecond laser pulses to drive the molecular transition in order to be insensitive to the rotational state distribution. Our implementation can be naturally extended to pump-probe experiments investigating intramolecular dynamics at ultrafast timescales~\cite{schindlerUltrafastInfraredSpectroscopy2019}. 
We also expect that this method can be a useful tool for high-precision spectroscopy. A high spectral resolution can be reached by a pulse train originating from a narrow linewidth laser with the repetition period being an integer multiple of the ions’ oscillation period and a pulse duration that is shorter than the ion’s oscillation period. This enables a coherent momentum kick that can be detected with the cat state while probing the transition at a linewidth that is limited by Fourier transform of the pulse train. This interaction is equivalent to a high-precision interrogation of a transition with a frequency comb which has been demonstrated in atoms and molecules~\cite{chouFrequencycombSpectroscopyPure2020}. Naturally, performing high-precision spectroscopy on complex molecules will require state preparation to reach an appreciable signal to noise ratio. We anticipate that recoil detection with spectrally shaped laser sources in combination with traditional quantum logic methods~\cite{chouPreparationCoherentManipulation2017} can provide powerful tools to enable such state preparation.

Our experiments demonstrate and characterize a method to detect single photon absorption events on a single molecular ion, which is a major step towards realizing non-destructive quantum state detection measurements in a wide class of molecular ions. 
With large enough amplification of the recoil signal, one can realize the detection of a single photon absorption event directly as an atomic excitation in a single shot. 
Furthermore, other methods to increase the detection sensitivity of the photon recoil exist~\cite{loSpinMotionEntanglement2015,burdQuantumAmplificationMechanical2019,bondOptimalDisplacementSensing2025} and might also be beneficial for realizing single-shot measurements. 
With selective single-shot measurement in an effective three-level system, the recoil detection can enable non-destructive measurements on two of the states~\cite{braginskyQuantumNondemolitionMeasurements1980,humeHighFidelityAdaptiveQubit2007}, as discussed in the Methods~\hyperref[meth:qnd]{F}. 
Such selective single-shot measurement will also be key for measurement-based molecular state preparation and readout, which are crucial ingredients in the development of quantum technologies with molecular ions~\cite{liuQuantumStateTracking2024,shlykovOptimizedStrategiesQuantumState2025,kimStateSelectiveIonizationTrapping2025}. 

The presented method can be applied to any molecular transition that produces a momentum kick of sufficient magnitude based on Equation~\ref{eq:ld param}. With our set of parameters, the method should be able to detect photon absorption events on transitions with a frequency down to 2000~\unit{\per\centi\metre}. The transitions considered in the discussion here are limited to vibrational transitions where the radiative lifetime of the excited states are much longer than time to create the cat state. If this is not the case, the momentum kick has also a diffusive part that reduces the achievable detection efficiency~\cite{hempelEntanglementenhancedDetectionSinglephoton2013}, which applies to dipole allowed transitions in the visible and UV part of the spectrum.

\section*{Acknowledgments}
This research was funded by Austrian science fund FWF (10.55776/COE1, 10.55776/PIN3213524), the Austrian Research Promotion Agency (FFG) under the project “HDCode” (FO999921407), and the ERC Horizon 2020 project ERC-2020-STG 948893. This research is funded in
part by the Gordon and Betty Moore Foundation through Grant GBMF12992. 
V.Š. was supported by the funding from the MEYS of Czech Republic under the project CZ.02.01.01/00/22 008/0004649.
M.G. and M.T. gratefully acknowledge the National Science Centre, Poland (grant
no.~2021/43/B/ST4/03326) for financial support and Poland's high-performance computing infrastructure PLGrid (HPC Centers: ACK Cyfronet AGH) for providing computer facilities and support (computational grant no.~PLG/2024/017527).
The authors acknowledge Artem Zhdanov, Zhimin Liu, Fabian Wolf, and Stefan Willitsch for discussions.

\section*{Author Contributions}
Z.W., B.F. and P.S. designed the experiment. Z.W. and T.D. carried out the measurement and analyzed the data. Z.W., M.I.M., B.F., M.K., R.N. and P.S. contributed to the experimental setup. Z.W., V.Š., B.F. and P.S. developed the model for explaining the result. M.G. and M.T. performed electronic structure calculations. Z.W., T.D., B.F., P.S., M.G. and M.T. contributed to the manuscript. All authors reviewed the manuscript. B.F., M.T. and P.S. supervised the project.

\section*{Conflict of Interest}
The authors have no conflicts of interest to disclose.

\section*{Methods}

\subsection{Molecule generation}
\label{meth:molecule generation}
Our experiment is performed on a mixed-species two-ion crystal consisting of $^{40}$Ca$^+$ and $^{40}$CaOH$^+$ confined in a linear Paul trap~\cite{wuPhotodissociationSpectraSingle2024a}. To form such a crystal, we first load two $^{40}$Ca$^+$ ions and then leak in water vapor with a leak valve from a gas chamber that contains mainly water vapor with a pressure of around 10$^{-6}$~mbar. In the process, we apply 397~nm laser that drives the $4^2S_{1/2} \rightarrow 4^2P_{1/2}$ dipole transition in $^{40}$Ca$^+$ to cool the ion crystal and monitor the fluorescence from $^{40}$Ca$^+$ ions with a camera. As the molecule does not fluoresce, the generation of a molecular ion leads to one of the ions turning dark on the camera. When this is observed, we switch off the leak valve and perform mass spectrometry on the dark ion by measuring the motional frequency of the ion crystal~\cite{homeQuantumScienceMetrology2013} to verify that the generation of CaOH$^+$. In general, it takes around 5 minutes to form a molecule. After closing the valve, the pressure in the vacuum chamber returns to around 10$^{-10}$~mbar within a few minutes. 

After generating the mixed-species ion crystal, background gas collisions occur, leading to the atomic and molecular ions swapping positions on the timescale of several seconds. For cat state spectroscopy measurements, we work with only one of the two ion crystal configurations since the two configurations exhibit different motional frequencies. This is likely due to imperfect micromotion compensation~\cite{homeQuantumScienceMetrology2013}. We detect the ion crystal configuration during Doppler cooling before the spectroscopy sequence and attempt to alter the configuration randomly if the wrong configuration is detected. This is done by displacing the ion crystal radially from its trapping position. We repeat this procedure until the ions return to the correct configuration. 

\subsection{Cat state engineering}
\label{meth:cat state engineer}
The non-classical motional state for probing photon recoil is generated by a bichromatic laser beam applied to the $^{40}$Ca$^+$ ion. The two frequency components of equal intensity in the bichromatic beam are detuned by $-\omega_z$ and $+\omega_z$ from the atomic quadrupole transition between $\lvert \uparrow \rangle$ and $\lvert \downarrow \rangle$, with $\omega_z$ being the in-phase motional frequency of the ion crystal~\cite{hempelEntanglementenhancedDetectionSinglephoton2013}. The Lamb-Dicke parameter of this quadrupole transition in the experiment is given by 
\begin{equation}
    \eta_\text{a} = \sqrt{\frac{\hbar}{2 M_\text{a} \omega_z}} k_z^\text{a} e_z^\text{a},
\end{equation}
where $M_\text{a}$ is the atomic mass, $k_z^\text{a}$ is the wavevector component of the quadrupole transition laser along the trap axis, and $e_z^\text{a}$ is the participation factor of the atomic ion in the in-phase mode.
In the Lamb-Dicke regime, the light-ion interaction Hamiltonian can be written as \cite{leibfriedQuantumDynamicsSingle2003}
\begin{equation}
    \hat{H}_{\text{int}}=\hbar \eta_\text{a} \frac{\Omega_0}{2}(\hat{\sigma}_{+} \hat{a} e^{i\phi_r}+ \hat{\sigma}_{+} \hat{a}^{\dagger} e^{i\phi_b}+ h.c.),
\end{equation}
where $\Omega_0$ is the Rabi frequency, $\hat{a}^{\dagger}(\hat{a})$ is the creation(annihilation) operator of the harmonic oscillator describing in-phase motion, and $\phi_r$($\phi_b$) is the phase of the frequency component with detuning $-\omega_z$($+\omega_z$) in the light field. This corresponds to a spin-dependent displacement characterized by $\phi_{\pm} = (\phi_{b} \pm \phi_{r})/2$, given by
\begin{multline}
	\hat{H}_{\text{int}} = \hbar \eta_\text{a} \frac{\Omega_0}{2} \left(\hat{\sigma}_{x} \cos{\phi_{+}} - \hat{\sigma}_{y} \sin{\phi_{+}}\right) \\
    \left[i \left(- \hat{a} + \hat{a}^{\dagger}\right) \sin{\phi_{-}} + \left(\hat{a} + \hat{a}^{\dagger}\right) \cos{\phi_{-}}\right].
\end{multline}
Considering $\phi_+ = 0$, the interaction can be expressed as
\begin{equation}
    \hat{H}_{\text{int}}=\hbar \eta_\text{a} \frac{\Omega_0}{2} \hat{\sigma}_x \left(\hat{a} e^{- i \phi_-} + \hat{a}^{\dagger} e^{i \phi_-}\right).
    \end{equation}
This leads to opposite motional displacements of $\hat{D}(\alpha)$ and $\hat{D}(-\alpha)$ with $\alpha = - i \eta_\text{a} \Omega_0 T e^{i \phi_-} /2$ for the two $\hat{\sigma}_x$ eigenstates $\lvert \pm \rangle$ for a given pulse duration $T$. 

The generated state is thus a superposition of two wavepackets with different atomic states shown in Equation~\ref{eq:cat state}. By applying the interaction with $\phi_+ = \pi$ for the same duration, the two wavepackets can be combined again with the inverted spin-dependent displacement. 
Momentum recoil due to photon absorption is timed to occur between the generation and reversal of the cat state. 
This leads to geometric phase difference accumulated between the two wavepackets, which is then detected by measuring the atomic state as shown in Equation~\ref{eq:final state}. 

\subsection{Experimental imperfections}
\label{meth:imperfections}
Experimental imperfections are described by a physically motivated model including atomic and motional decoherence, which, in our case, result from fluctuations of the magnetic field and trap voltages respectively. We simulate the cat-state photon absorption detection using Lindblad master equations where we include the two decoherence channels as collapse operators:
\begin{equation}
    \hat{C}_\text{spin}=\sqrt{\frac{1}{2 T_\text{spin}}}\hat{\sigma}_z, \; \hat{C}_\text{motion}=\sqrt{\frac{2}{T_\text{motion}}}\hat{a}^\dagger \hat{a}
\end{equation}
where $T_\text{spin}$ is the coherence time of atomic two-level system and $T_\text{motion}$ is the coherence time of a superposition of the motional ground state and the first excited state. We estimated $T_\text{spin}=2.8(3)$~ms and $T_\text{motion}=85^{+98}_{-35}$~ms with Ramsey experiments at the time scale of the measurement. The motional coherence time is much longer than the duration of a cat-spectroscopy experiment which causes the large uncertainty on the estimated value. 
With these values, we can predict the signal produced by the experiment with various cat state amplitudes and various kick sizes, as shown in Figure~\ref{fig:fig3}. The purple shaded region indicates the simulation results considering one standard deviations of the atomic and motional coherence of the system. The dominating effect of these imperfection is a loss in signal amplitude, which we can also model with a simple heuristic model in Equation~\ref{eq:exp signal model} that multiplies the ideal contrast in Equation~\ref{eq:signal model} with a factor $\mathcal{S}_\text{max}$. For a cat state amplitude of $|\alpha|=6.5(7)$, we determine the value of $\mathcal{S}_\text{max}$ to be 0.52(3) by fitting Equation~\ref{eq:exp signal model} to the experimental data of various kick sizes. 

The value of $\mathcal{S}_\text{max}$ shows the effect of decoherence and varies for different cat state amplitudes and different durations of the measurement sequence. When investigating the effective photon absorption probability as a function of the number of femtosecond laser pulse $n_\text{pulse}$, as presented in Figure~\ref{fig:fig4}(a), the variation in $n_\text{pulse}$ leads to the change in the time interval between the generation and the reversal of the cat state. This results in the variation in $\mathcal{S}_\text{max}$ as a function of $n_\text{pulse}$. We investigated this effect with the electric kick measurement and model it as
\begin{equation}
    \mathcal{S}_\text{max}(t) = \mathcal{S}_\text{max} e^{-t/\tau_d},
\end{equation}
for $|\alpha|=6.5(7)$ where $\tau_d = 0.92(6)$~ms.
We thus adjust the measured signals for different numbers of laser pulses accordingly.

Another potential source of imperfections besides decoherence is the timing jitter of the photon absorption events with respect to the ion crystal motion. The magnitude of the recoil-induced geometric phase difference $\Phi=2\eta_m \text{Re}(\alpha)$ depends on the phase difference between the cat state and the displacement operator from the photon absorption recoil. Therefore, it is influenced by the free evolution time between the generation of the cat state and the photon absorption event $T_\text{wait}$. In our case, the photon absorption event is synchronized with the femtosecond laser pulses. In order to achieve a deterministic timing, we trigger the cat state generation sequence on the periodic femtosecond laser pulses with a timing jitter of less than 10~ns. This jitter is considerably shorter than the oscillation period of the ions and ensures a constant free evolution time $T_\text{wait}$ during the measurement. Additionally, the values of the trap frequency $\omega_z$ and the repetition rate of the laser $f_\text{rep}$ influence the phase. We set the trap frequency to be an integer multiple of the repetition rate $\omega_z/2\pi=4 \cdot f_\text{rep}$, so that we expect the same $\text{Re}(\alpha)$ for photon recoil from each of the femtosecond laser pulses. A variation of the trap frequency or the repetition rate, therefore, results in a varying phase for the signal obtained from each pulse. The variation of $\omega_z/2\pi$ is measured to be around 50~Hz and the fluctuation of $f_\text{rep}$ is identified to be below 10~Hz. The variation in $\arg(\alpha)$ is thus less than $3.2 \times10^{-2}$ for the maximum $T_\text{wait}$ and can be neglected.

\subsection{Quantum-chemical calculations}
\label{meth:quantum-chemical calculations}
The vibrational transition frequencies and intensities are calculated ab initio using state-of-the-art quantum-chemical methods. We focus on the observed O--H stretching mode, which forms the most intense spectral band. Initially, an accurate potential energy surface is computed using electronic structure coupled-cluster methods, from which the harmonic frequencies and normal coordinates of the stretching modes are obtained. The resulting best estimate for the O--H harmonic frequency is 3950~cm$^{-1}$. Subsequently, we neglect couplings between different normal modes, reducing the multidimensional anharmonic vibrational problem to a set of one-dimensional anharmonic vibrational problems along the normal coordinates. We solve the one-dimensional Schr\"odinger equation for nuclear motion along the normal coordinate dominated by the O--H stretching using the discrete variable representation method~\cite{tiesingaPhotoassociativeSpectroscopyHighly1998}. The computed vibrational energies and wavefunctions give the fundamental transition frequency of 3792~cm$^{-1}$ and oscillator strength of $3.7 \times 10^{-5}$. Finally, couplings of the fundamental O--H stretching mode with the Ca--O stretching and Ca--O--H bending modes are reintroduced using vibrational second-order perturbation theory (VPT2)~\cite{bloinoVPT2RouteNearInfrared2015}, which lowers the O--H fundamental frequency by 9.5~cm$^{-1}$ and results in its final value of 3783~cm$^{-1}$.

The electronic structure computations employ the range of correlation-consistent polarized weighted core-valence orbital basis sets augmented with diffuse functions (aug-cc-pwCV$n$Z-PP, $n$ = T, Q, 5)~\cite{hillGaussianBasisSets2017,petersonAccurateCorrelationConsistent2002,kendallElectronAffinitiesFirstrow1992}. The ten core electrons of calcium are replaced by the ECP10MDF effective core potential to account for relativistic effects~\cite{limRelativisticSmallcoreEnergyconsistent2006}. The potential energy surface is obtained using a composite scheme in which the total energy is expressed as the sum of three contributions: (i) the Hartree-Fock mean-field energy, (ii) the leading part of the correlation energy obtained using the coupled-cluster method restricted to single, double, and noniterative triple excitations [CCSD(T)]~\cite{stantonWhyCCSDTWorks1997}, and (iii) the contribution from triple excitations in CCSDT neglected by CCSD(T)~\cite{nogaFullCCSDTModel1987}. The latter term is computed using a triple-$\zeta$ basis set, whereas the first two terms are extrapolated to the complete basis set (CBS) limit employing consecutive basis sets up to quintuple-$\zeta$. The three-point exponential~\cite{halkierBasissetConvergenceEnergy1999} and two-point $1/n^3$~\cite{helgakerBasissetConvergenceCorrelated1997} extrapolation formulas are used for the Hartree-Fock and correlation energies, respectively.  The permanent electric dipole moment is calculated using the analytical-derivative technique at the CCSD(T)/aug-cc-pwCV5Z level. Its value for the ground vibrational level is 6.2~D. 

The potential energy surface of linear CaOH$^+$ is computed as a function of the Ca--O and O--H bond lengths on a uniform two-dimensional grid with a spacing of 0.002~bohr around the equilibrium geometry~\cite{wuPhotodissociationSpectraSingle2024a} and interpolated with a natural cubic spline. After determining the normal coordinate, thirty-three electronic energies are calculated along this coordinate. The VPT2 correction is estimated from anharmonic constants obtained at the CCSD(T)/aug-cc-pwCVTZ level.

All electronic structure and VPT2 calculations are performed with the \textsc{CFOUR} v2.1 program~\cite{matthewsCoupledclusterTechniquesComputational2020}.

\subsection{Molecular excitation dynamics}
\label{meth:molecular excitation dynamics}
We model the excitation dynamics of the O--H stretching vibrational transition of CaOH$^+$ in the experiment as follows. The molecule is considered here as a two-level system consisting of the ground and first excited states, $\lvert g \rangle$ and $\lvert e \rangle$. The intensity of the laser pulses applied to the molecule is assumed to have a Gaussian temporal profile with a full width at half maximum (FWHM) pulse duration of $\tau=2\tau_\sigma\sqrt{2\ln2}$, where $\tau_\sigma$ is the 1-sigma pulse duration. The molecule then experiences the following time-dependent electric field
\begin{equation}
    \vec{E}(t) = \vec{E}_0  e^{-t^2/4\tau_\sigma^2} (e^{-i(\omega t - \phi)} + \text{c.c.}),
\end{equation}
where $\omega$ and $\phi$ are the center frequency and the phase of the laser and $E_0$, the peak electric field strength, can be calculated from the average laser intensity $I_\text{avg}$ as $E_0 = \sqrt{I_\text{avg} /( f_\text{rep} \tau \sqrt{\pi/\ln2} \cdot \epsilon_0 c)}$.
The light-molecule interaction can then be described as
\begin{equation}
    \hat{H}_\text{m}(t) = - \vec{d} \cdot \vec{E}(t) = \mu_{eg} E(t) \cos{\theta},
\end{equation}
where the transition dipole moment $\mu_{eg}$ can be related to the oscillator strength as $\mu_{eg}=\sqrt{f_{eg} \cdot 3\hbar e^2/m_e\omega_0}$ and $\theta$ is the angle between the molecular axis and the direction of the polarization of the light field. In the interaction picture under the rotating-wave approximation, the interaction can be described by
\begin{equation}
    \label{eq:pulse hamiltonian}
    \hat{H}_\text{m}^\text{I}(t) = \hbar \frac{\Omega(t)}{2} \left( \lvert e \rangle \langle g \rvert e^{i\phi} + \text{h.c.} \right) + \hbar \Delta \lvert g \rangle \langle g \rvert,
\end{equation}
where $\Delta=\omega - \omega_0$ is the detuning of the laser from the transition frequency $\omega_0$ and $\Omega(t)$ is the Rabi rate given by
\begin{equation}
    \Omega(t) = \frac{2\mu_{eg}E_0 \cos\theta}{\hbar} e^{-t^2/4\tau_\sigma^2}.
\end{equation}

We aim to describe the experimental results using this model with no free parameters with a Monte-Carlo simulation of the light-molecule interaction expressed in Equation~\ref{eq:pulse hamiltonian}. In each trial of the simulation, for each laser pulse in the train, we randomly assign the angle $\theta$, assuming that the molecule is randomly oriented in the experiment. We also assign the phase $\phi$ of the light field with respect to the molecular transition randomly, considering the lack of phase stability of the laser from pulse to pulse and the dephasing due to the molecule's mixed rotational state. The Rabi rate of a single laser pulse is calculated based on ab initio calculations on the transition frequency and oscillator strength (see Methods~\hyperref[meth:quantum-chemical calculations]{D}) and the laser parameters described below.

In our setup, the femtosecond laser pulses are produced by an ORPHEUS-HP optical parametric amplifier (OPA) pumped by a CARBIDE-CB5 femtosecond laser, both produced by Light Conversion. 
The light applied to the ions is linearly polarized in the vertical plane and with our magnetic field geometry this results in an equal superposition of $\sigma_+$ and $\sigma_-$ polarizations. The average intensity at the position of the ions is $1.1(1) \times 10^{4}$~\unit{\watt\per\square\centi\metre}.
The spectrum of the light field is measured with a Redstone OSA305 optical spectrum analyzer to determine the center wavelength and linewidth for each frequency measured in the presented absorption spectrum. We observe absorption lines due to the presence of water vapor in the beam path, but we assume that this is not affecting the absorption spectroscopy experiment due to power broadening.

The measurements presented in Figure~\ref{fig:fig4}(a) and Figure~\ref{fig:fig4}(c) are performed with a laser center frequency of $\nu=3703.3(2)$~\unit{\per\centi\metre} and a FWHM linewidth of \SI{126.7(4)}{\per\centi\metre}. The simulation is based on these parameters and the assumption that the laser pulses are Fourier transform-limited with a FWHM pulse duration of 116.2(4)~fs. 
For the absorption spectrum shown in Figure~\ref{fig:fig4}(d), the experimental results are compared to the simulation performed with the Fourier transform-limited pulse durations given by the measured center frequency and linewidth at each of the OPA wavelength settings. 
 
\subsection{Non-destructive state detection}
\label{meth:qnd}

Photon absorption detection can provide an effective non-destructive state detection acting on a specific subspace of the molecular degrees of freedom. We sketch an example of such a measurement method, acting on the two lowest states of a vibrational mode in a molecule. Consider the three lowest energy vibrational states $\lvert 0 \rangle , \, \lvert 1 \rangle , \,\lvert 2 \rangle $. We assume and that we have the capability to perform selective operations on these states which can be spectroscopically addressed because the molecule's anharmonicity  yields unique transition frequencies. The proposed technique provides a projective measurement in the $\{ \lvert 0 \rangle , \, \lvert 1 \rangle \}$ manifold. We assume that the system is in a superposition of the vibrational states, the shared motion and the atom are in their ground state \[ (a \lvert 0 \rangle + b \lvert 1 \rangle ) \otimes \lvert \downarrow \rangle \otimes \lvert 0 \rangle_z  \, .\] We will then create the cat state using the atom-motion interaction, yielding the atom-motion state \[ \lvert \Psi \rangle_\text{cat} = \frac{1}{\sqrt{2}} \lvert + \rangle \lvert\alpha\rangle_z + \lvert - \rangle \lvert-\alpha\rangle_z \, .\] We assume that the cat state spectroscopy is set up to completely flip the atomic electronic state if a photon absorption has occurred $\text{Re}(\alpha) = \pi/(4 \eta_\text{m})$. The combined state of the molecular, atomic, and motional degrees of freedom is then
\[
( a \lvert 0 \rangle + b \lvert 1 \rangle ) \otimes \lvert \Psi \rangle_\text{cat} \, .
\]
We then assume that we can perfectly transfer the population from the $\lvert 1 \rangle$ state to the state $\lvert 2 \rangle$ with a laser pulse which applies a displacement operator on the ions' motion
\[
 a \lvert 0 \rangle \otimes  \lvert \Psi \rangle_\text{cat} + b \lvert 2 \rangle  \otimes \hat{D}(\eta_\text{m}) \lvert \Psi \rangle_\text{cat}  \, .
\]
After reversing the cate state generation, the state is then
\[
 \biggl(a \lvert 0 \rangle  \otimes \lvert \downarrow \rangle  + b \lvert 2 \rangle  \otimes \lvert \uparrow \rangle  \biggr) \otimes \lvert 0 \rangle_z \, ,
\]
representing an entangled state between the molecular vibrational degree of freedom and the electronic state of the atom. A projective measurement on the atomic electronic state will thus also project the molecular state in either $\lvert 0 \rangle$ or $\lvert 2 \rangle$. If the state is projected into $\lvert 2 \rangle$, the population can be brought back to $\lvert 1 \rangle$ which concludes the method and provides a non-destructive state detection.

\subsection*{Data availability}
The data that support the findings of this study are openly available at \href{https://doi.org/10.5281/zenodo.17735414}{10.5281/zenodo.17735414}.

\bibliography{CatSpec202508}

\end{document}